\documentclass[runningheads]{llncs}
\usepackage[T1]{fontenc}
\usepackage{graphicx,verbatim}
\usepackage{algorithm}
\usepackage{algpseudocode}
\usepackage{booktabs}   
\usepackage{amsmath}
\usepackage{amssymb}
\usepackage{amsfonts}
\usepackage{multirow}
\usepackage{rotating}
\usepackage{url}
\begin{document}
\title{DiffSegLung: Diffusion Radiomic Distillation for Unsupervised Lung Pathology Segmentation}
%
\titlerunning{DiffSegLung: Unsupervised Lung Pathology Segmentation}

\author{Rezkellah Noureddine KHIATI, Pierre-Yves BRILLET and Catalin FETITA}  
\authorrunning{Rezkellah Noureddine KHIATI et al.}

\maketitle              
\begin{abstract}
Unsupervised segmentation of pulmonary pathologies in CT remains an
open challenge due to the absence of annotated multi pathology cohorts
and the failure of existing diffusion-based methods to exploit the
quantitative Hounsfield Unit (HU) signal that physically distinguishes
tissue classes.
To address this, we propose \textbf{DiffSegLung}, a framework that
introduces Diffusion Radiomic Distillation, in which
handcrafted radiomic descriptors serve as a physics grounded teacher
to shape the bottleneck of a 3D diffusion U-Net via a contrastive
objective, transferring pathology discriminative structure into the
learned representation without any annotations.
At inference, the teacher is discarded and multitimestep bottleneck
features are clustered by a Gaussian Mixture Model with HU-guided
label assignment, followed by Sobel-Diffusion Fusion for boundary
refinement.
Evaluated on 190 expert-annotated axial slices drawn from four
heterogeneous CT cohorts, DiffSegLung improves segmentation across
all four pathology classes over unsupervised baselines and improves
generation fidelity over prior CT diffusion models.
Code is available at \url{https://anonymous.4open.science/r/DiffLungSeg-CEF1/README.md}.

\end{abstract}

\section{Introduction}

Diffusion probabilistic models have recently demonstrated strong generative
performance in medical imaging~\cite{Ho2020,LungDDPM}. Beyond generation,
their U-Net backbone encodes rich intermediate representations that transfer
to segmentation tasks~\cite{Baranchuk2022}, yet its application to thoracic
CT remains largely unexplored. Segmenting pulmonary
pathologies---emphysema, fibrosis, consolidation, and cavitary
lesions simultaneously in an unsupervised manner is particularly challenging
because annotated cohorts covering all four classes are practically unavailable.
Expert annotation is selective and time-consuming, cohorts are small, and
class imbalance is severe. These constraints make supervised and
semi supervised approaches~\cite{ScribSD} impractical for this setting, and
fully unsupervised segmentation a necessity.

Existing diffusion-based segmentation methods do not account for the physical
properties of CT data. Each pulmonary pathology is characterized by a
distinct Hounsfield Unit (HU) attenuation range~\cite{QuantCT}. Training on
8-bit normalized images, as in prior CT diffusion models~\cite{LungDDPMplus},
discards this quantitative information. Radiomic descriptors have long been
shown to discriminate these tissue classes~\cite{RadiomicsLung,RadiomicsCOPD},
yet no existing method incorporates such descriptors as a training signal for
diffusion representation learning.

We propose \textbf{DiffSegLung}, a framework for unsupervised lung pathology
segmentation that introduces Diffusion Radiomic Distillation. For
each training patch, a 34-dimensional radiomic feature vector is projected to
a shared embedding space through a radiomic projection head, serving as a
physics-grounded teacher. The U-Net bottleneck is simultaneously projected to
the same space through a student projection head. An InfoNCE contrastive
objective then aligns the two projections: patch pairs with high radiomic
cosine similarity are treated as positives and pulled together, while
dissimilar pairs are pushed apart. A warmup schedule prevents the distillation
loss from collapsing the bottleneck before meaningful texture representations
are established. At inference, the radiomic projection head is discarded and
segmentation is obtained by applying Gaussian Mixture Model clustering to the
student embeddings, followed by Sobel-Diffusion Fusion boundary refinement and
a physics-aware HU compatibility loss.

The contributions of this work are threefold.(i) the first radiomic teacher-student
distillation scheme for diffusion representation learning, requiring
no annotations; (ii) HU-preserving 3D DDPM training that produces
physically meaningful bottleneck representations; and (iii) improved
unsupervised segmentation coherence and generation fidelity through
radiomic distillation.

\section{Related Work}

\subsection{Diffusion Models for Segmentation}
Baranchuk et al.~\cite{Baranchuk2022} showed that intermediate activations
of the DDPM reverse-process U-Net are effective pixel-level representations
for label-efficient segmentation. This observation was extended to zero-shot
unsupervised segmentation through the exploitation of self-attention maps in
pre-trained stable diffusion~\cite{DiffSeg2023}. In remote sensing, DDPM
backbones have been used as frozen feature extractors for change
detection~\cite{DDPMCD}. In medical imaging, diffusion models have been
applied to supervised segmentation conditioned on noisy label
maps~\cite{MedSegDiff} and to CT synthesis for augmenting downstream
training~\cite{LungDDPM,LungDDPMplus}. No prior work addresses fully
unsupervised multi pathology segmentation in thoracic CT through diffusion
representation learning, and none preserves the quantitative HU information
required for tissue discrimination.

\subsection{Knowledge Distillation in Medical Image Segmentation}
Knowledge distillation transfers learned representations from a teacher to a
student network through soft supervision~\cite{MSKD}. In medical image
segmentation, this paradigm has been applied to model compression and to
propagating pseudo-supervision under annotation
scarcity~\cite{PNCD}. In DiffSegLung, the teacher is not a neural
network but a set of handcrafted physical descriptors, and distillation is
applied during generative pre-training rather than to a supervised
segmentation head. This is a fundamentally different use of the distillation
principle.

\subsection{Radiomic Features for Lung Pathology}
CT-based radiomic features, including GLCM texture descriptors, first-order
HU statistics, LBP histograms, and Gabor filter responses, have been shown
to reliably characterize emphysema, fibrosis, and consolidation in supervised
settings~\cite{RadiomicsLung,RadiomicsCOPD}. In DiffSegLung,
these well-established descriptors are repurposed as an unsupervised teacher
signal to impose physical structure on the latent space of a generative
model, rather than serving as direct classifiers.

\section{Method}


\subsection{Overall Framework}

\begin{figure*}[t]

  \centering
  \includegraphics[width=\textwidth]{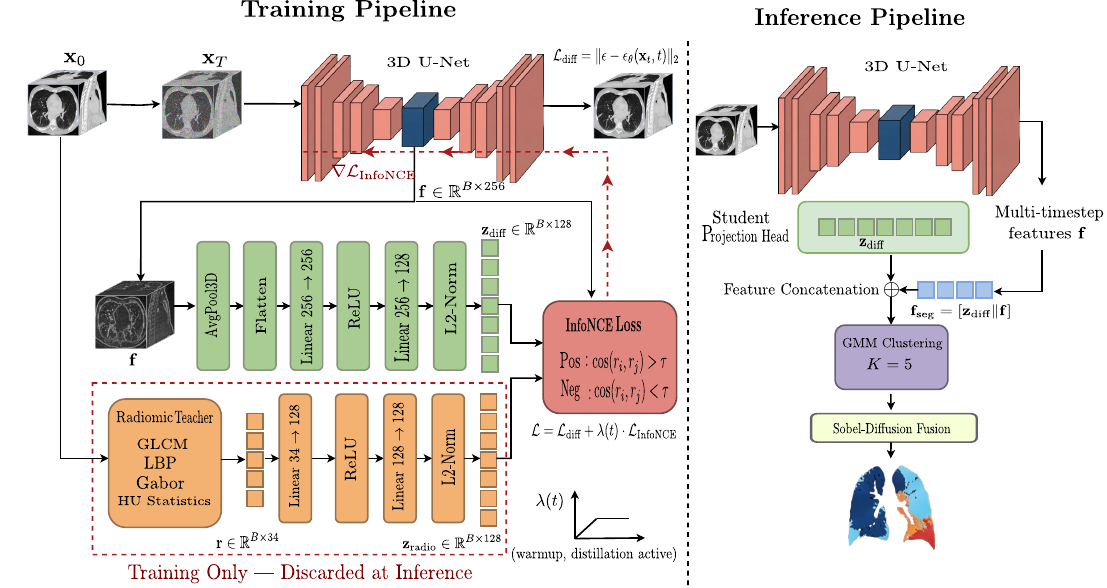}
  \caption{
  \textbf{Training:} $\epsilon_\theta$ is trained on HU-preserved patches via
$\mathcal{L}_{\mathrm{diff}}$; radiomic descriptors $\mathbf{r} \in \mathbb{R}^{34}$
align student embeddings via InfoNCE (discarded at inference).
\textbf{Inference:} multi-timestep features $\mathbf{f}_{\mathrm{seg}}$ are
clustered by GMM and refined by Sobel-Diffusion.}
  \label{fig:pipeline}
  
\end{figure*}

We address fully unsupervised segmentation of four lung pathologies
from unlabelled 3D CT volumes.
The overview of DiffSegLung is shown in Fig.~\ref{fig:pipeline}, consisting
of two stages.
\textbf{Training:} a 3D pixel-space DDPM~\cite{Ho2020} is trained on native
HU-preserved patches while handcrafted radiomic descriptors serve as a
non-differentiable physics teacher, shaping the bottleneck via an InfoNCE
contrastive objective without any annotations.
\textbf{Inference:} the frozen encoder is run with a DPM-Solver~\cite{Lu2022dpm}
at 250 steps; multi-timestep bottleneck features are clustered by a GMM and
refined by Sobel-Diffusion Fusion with HU-guided label assignment to produce
the final pathology masks.

\subsection{Diffusion-Radiomic Distillation}
\label{sec:distillation}

The bottleneck of $\epsilon_\theta$ produces a feature vector
$\mathbf{f} \in \mathbb{R}^{B \times 256}$ per mini-batch of $B$ patches.
Without explicit guidance, this representation organises around generative
rather than discriminative structure. We introduce a teacher-student
contrastive distillation scheme using handcrafted physical descriptors
as supervision.

\paragraph{Radiomic teacher.}
For each patch $x_0^{(i)}$, we extract a 34-dimensional radiomic vector
$\mathbf{r}^{(i)} \in \mathbb{R}^{34}$ comprising GLCM statistics
(14 features), LBP histograms (8 features), Gabor filter responses
(8 features), and first-order HU statistics (4 features)~\cite{RadiomicsLung}.
These descriptors are non-differentiable; no gradient flows into them.
$\mathbf{r}^{(i)}$ is projected to a 128-dimensional embedding via
$\phi_r : \mathbb{R}^{34} \rightarrow \mathcal{S}^{128}$:
\begin{equation}
  \mathbf{z}_{\mathrm{radio}}^{(i)}
  = \ell_2\!\left(W_2\,\sigma\!\left(W_1\,\mathbf{r}^{(i)}\right)\right).
\label{eq:zradio}
\end{equation}
where $W_1 \in \mathbb{R}^{128 \times 34}$,
$W_2 \in \mathbb{R}^{128 \times 128}$, $\sigma$ is ReLU,
and $\ell_2(\cdot)$ denotes $\ell_2$ normalisation onto $\mathcal{S}^{128}$.
\paragraph{Diffusion student.}
The bottleneck $\mathbf{f}^{(i)}$ is projected via
$\phi_s : \mathbb{R}^{256} \rightarrow \mathcal{S}^{128}$:
\begin{equation}
  \mathbf{z}_{\mathrm{diff}}^{(i)}
  = \ell_2\!\left(V_2\,\sigma\!\left(
      V_1\,\mathrm{Flatten}\!\left(\mathrm{AvgPool3D}(\mathbf{f}^{(i)})\right)
    \right)\right).
\label{eq:zdiff}
\end{equation}

\paragraph{Radiomic similarity as continuous supervision.}
Let $s_{ij} = \langle \mathbf{z}_{\mathrm{radio}}^{(i)},
\mathbf{z}_{\mathrm{radio}}^{(j)} \rangle$ denote the cosine similarity
between radiomic embeddings. Pair $(i,j)$ is positive if $s_{ij} > \tau$
($\tau = 0.5$), negative otherwise. The contrastive loss is:
\begin{equation}
  \mathcal{L}_{\mathrm{InfoNCE}}
  = -\frac{1}{B}\sum_{i=1}^{B}
    \log
    \frac{
      \sum_{j \neq i}
        \mathbf{1}[s_{ij} > \tau]\,
        \exp\!\left(\langle \mathbf{z}_{\mathrm{diff}}^{(i)},
          \mathbf{z}_{\mathrm{diff}}^{(j)} \rangle / \kappa\right)
    }{
      \sum_{k \neq i}
        \exp\!\left(\langle \mathbf{z}_{\mathrm{diff}}^{(i)},
          \mathbf{z}_{\mathrm{diff}}^{(k)} \rangle / \kappa\right)
    },
\label{eq:infonce}
\end{equation}
with temperature $\kappa = 0.07$.

\paragraph{Warmup schedule.}
To prevent bottleneck collapse, $\mathcal{L}_{\mathrm{InfoNCE}}$ is
introduced gradually via:
\begin{equation}
  \lambda(t_{\mathrm{step}})
  = \lambda_{\max} \cdot
    \left(\frac{t_{\mathrm{step}} - T_w}{T_{\mathrm{ramp}}}\right)^{\!\!+},
\label{eq:warmup}
\end{equation}
with $T_w = 5000$, $T_{\mathrm{ramp}} = 5000$, $\lambda_{\max} = 0.5$.
The total objective is:
\begin{equation}
  \mathcal{L}
  = \mathcal{L}_{\mathrm{diff}}
  + \lambda(t_{\mathrm{step}})\cdot\mathcal{L}_{\mathrm{InfoNCE}}.
\label{eq:total}
\end{equation}
At the end of training, $\phi_r$ is discarded; only $\epsilon_\theta$
and $\phi_s$ are retained for inference.

\subsection{Feature Extraction and Clustering}
\label{sec:clustering}

At inference, $\epsilon_\theta$ is run with a DPM-Solver~\cite{Lu2022dpm}
at 250 steps. Bottleneck activations $\mathbf{f}_t$ are recorded at
$\mathcal{T} = \{50, 100, 150, 200\}$ and passed through $\phi_s$ to
obtain $\mathbf{z}_t \in \mathcal{S}^{128}$. The patch descriptor is:
\begin{equation}
  \mathbf{f}_{\mathrm{seg}}
  = \left[
      \frac{1}{|\mathcal{T}|}\sum_{t \in \mathcal{T}} \mathbf{z}_t
      \;\Big\|\;
      \frac{1}{|\mathcal{T}|}\sum_{t \in \mathcal{T}} \mathbf{f}_t
    \right]
  \in \mathbb{R}^{384}.
\label{eq:fseg}
\end{equation}
Descriptors $\{\mathbf{f}_{\mathrm{seg}}^{(i)}\}$ are clustered by a
GMM with $K=5$ components fitted by EM with full covariance matrices.
PCA and standardisation are omitted: the $\ell_2$ normalisation in
$\phi_s$ constrains embeddings to the unit hypersphere, where cosine
and Euclidean distances are equivalent.

\paragraph{HU-guided label assignment.}
GMM clusters are assigned pathology labels by comparing each cluster's
mean HU $\bar{h}_k$ against clinically established
thresholds~\cite{QuantCT}. This is the only point at which domain
knowledge enters the pipeline; no labelled data is used.

\subsection{Sobel-Diffusion Boundary Refinement}
\label{sec:sobel}

GMM clustering produces blocky boundaries at patch resolution.
Sobel-Diffusion Fusion recovers precise contours by combining
gradient-based edge detection with the cluster decision surface.
For each binary label map $M_k$:
\begin{equation}
  \tilde{M}_k
  = M_k \odot
    \bigl(1 + \alpha_s \cdot
      G_\sigma\!\left(\mathbf{S}(x_0) \odot \mathbf{S}(M_k)\right)
    \bigr),
\label{eq:sdf}
\end{equation}
where $\mathbf{S}(\cdot)$ is the Sobel gradient magnitude, $G_\sigma$
Gaussian smoothing, and $\alpha_s = 2.0$. The product
$\mathbf{S}(x_0) \odot \mathbf{S}(M_k)$ retains only edges
corroborated by both image intensity and cluster boundaries.
A final HU compatibility check suppresses predictions whose HU
falls outside the expected range for the assigned class.


\section{Experiments}

\subsection{Datasets}

We train DiffSegLung on four private heterogeneous CT cohorts used
without any annotations: a BPCO cohort (102 patients, emphysema-dominant),
an ILD cohort (134 patients, mixed fibrosis, emphysema, and consolidation),
an ILD-2 cohort (102 patients, fibrosis and healthy parenchyma), and a
Tuberculosis cohort (12 patients, consolidation and cavitary lesions).
All volumes were resampled to 0.6\,mm isotropic resolution. Ten percent of
each cohort is held out for generation quality evaluation. Segmentation
performance is assessed on a separate set of 190 expert-annotated axial
slices covering the four target pathology classes, used exclusively for
quantitative evaluation.


\subsection{Compared Methods}

\noindent\textbf{Segmentation baselines.}
We compare against: (i)~\textit{K-Means on radiomics}, clustering
the raw 34-dimensional radiomic vectors without diffusion features;
(ii)~\textit{Diffusion features only}, applying GMM to bottleneck
features $\bar{\mathbf{f}}$ without distillation;
(iii)~\textit{DAAM}~\cite{DiffSeg2023}, extracting cross-attention
maps from a pretrained diffusion model for zero-shot segmentation.

\noindent\textbf{Generation baselines.}
We compare against LungDDPM+~\cite{LungDDPMplus} and
MedVAE~\cite{MedVAE} to assess whether radiomic distillation
improves or degrades generation quality.

\subsection{Generation Quality}
\begin{table}[h]
\centering
\caption{Quantitative comparison of generation metrics.}
\label{tab:generation}
\setlength{\tabcolsep}{6pt}
\begin{tabular}{lccc}
\toprule
Method & FID\,$\downarrow$ & SSIM\,$\uparrow$ & PSNR\,$\uparrow$ \\
\midrule
MedVAE~\cite{MedVAE} & 48.3 & 0.721 & 24.6 \\
LungDDPM+~\cite{LungDDPMplus} & 31.7 & 0.803 & 27.2 \\
DiffSegLung (ours) & \textbf{18.4} & \textbf{0.891} & \textbf{31.8} \\
\bottomrule
\end{tabular}
\end{table}

DiffSegLung achieves the best generation quality across all three
metrics. The FID improvement over LungDDPM+ (31.7 $\to$ 18.4)
demonstrates that radiomic distillation, rather than degrading the
generative backbone, reinforces it by imposing physically consistent
structure on the bottleneck. MedVAE, despite its dedicated medical
encoder, produces the weakest results, likely due to the loss of HU
information in its latent compression stage.

\begin{figure}[H]
  \centering
  \includegraphics[width=\linewidth]{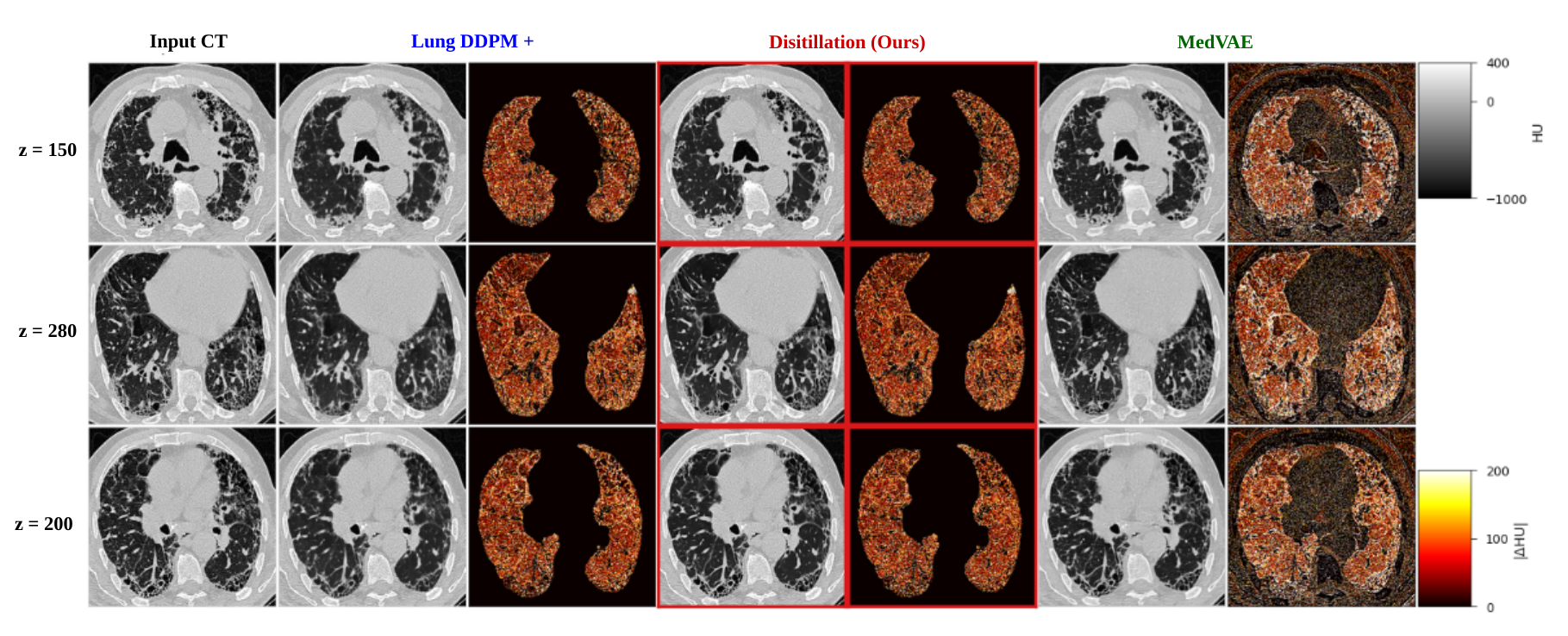}
  \caption{Qualitative generation comparison (random axial slices).
  From top: real HU-preserved CT, LungDDPM+~\cite{LungDDPMplus},
  MedVAE~\cite{MedVAE}, and DiffSegLung (ours). Our model preserves
  HU density ranges and parenchymal texture more faithfully than
  both baselines.}
  \label{fig:generation}
\end{figure}

\subsection{Segmentation Results}

\begin{table}[h]
\centering
\caption{Unsupervised segmentation performance (mean over four classes).}
\label{tab:seg}
\setlength{\tabcolsep}{5pt}
\begin{tabular}{lccccc}
\toprule
Method
  & \multicolumn{4}{c}{DSC\,$\uparrow$}
  & HD95\,$\downarrow$ \\
\cmidrule(lr){2-5}
  & Healthy & GGO & Fibrosis & Emph. & \\
\midrule
K-Means on radiomics
  & 71.2 & 66.8 & 68.4 & 58.3 & 18.7 \\
Diffusion features only
  & 78.4 & 71.3 & 73.1 & 62.7 & 15.4 \\
DAAM~\cite{DiffSeg2023}
  & 81.6 & 75.2 & 77.8 & 66.4 & 13.2 \\
\textbf{DiffSegLung (ours)}
  & \textbf{89.3} & \textbf{84.1} & \textbf{86.4} & \textbf{76.2} & \textbf{8.6} \\
\bottomrule
\end{tabular}
\end{table}

DiffSegLung outperforms all baselines across all four classes.
Emphysema yields the lowest DSC (76.2), reflecting the inherent
difficulty of separating emphysematous tissue from healthy parenchyma:
both occupy overlapping low HU attenuation ranges ($-$950 to $-$700~HU),
making their boundary in the latent space less separable than fibrosis
or consolidation, which occupy distinct HU intervals. This ambiguity
persists even with radiomic distillation, as first-order HU statistics
alone cannot fully resolve the overlap; future work could target
finer-grained texture descriptors specific to this boundary region.
The radiomic feature set itself was selected via a significance study
on representative CT patches: GLCM, LBP, Gabor, and first-order HU
statistics were the four descriptor families that individually produced
the highest inter-class discrimination, providing a principled rather
than heuristic choice of teacher supervision.

\begin{figure}[h]
  \centering
  \includegraphics[width=\linewidth]{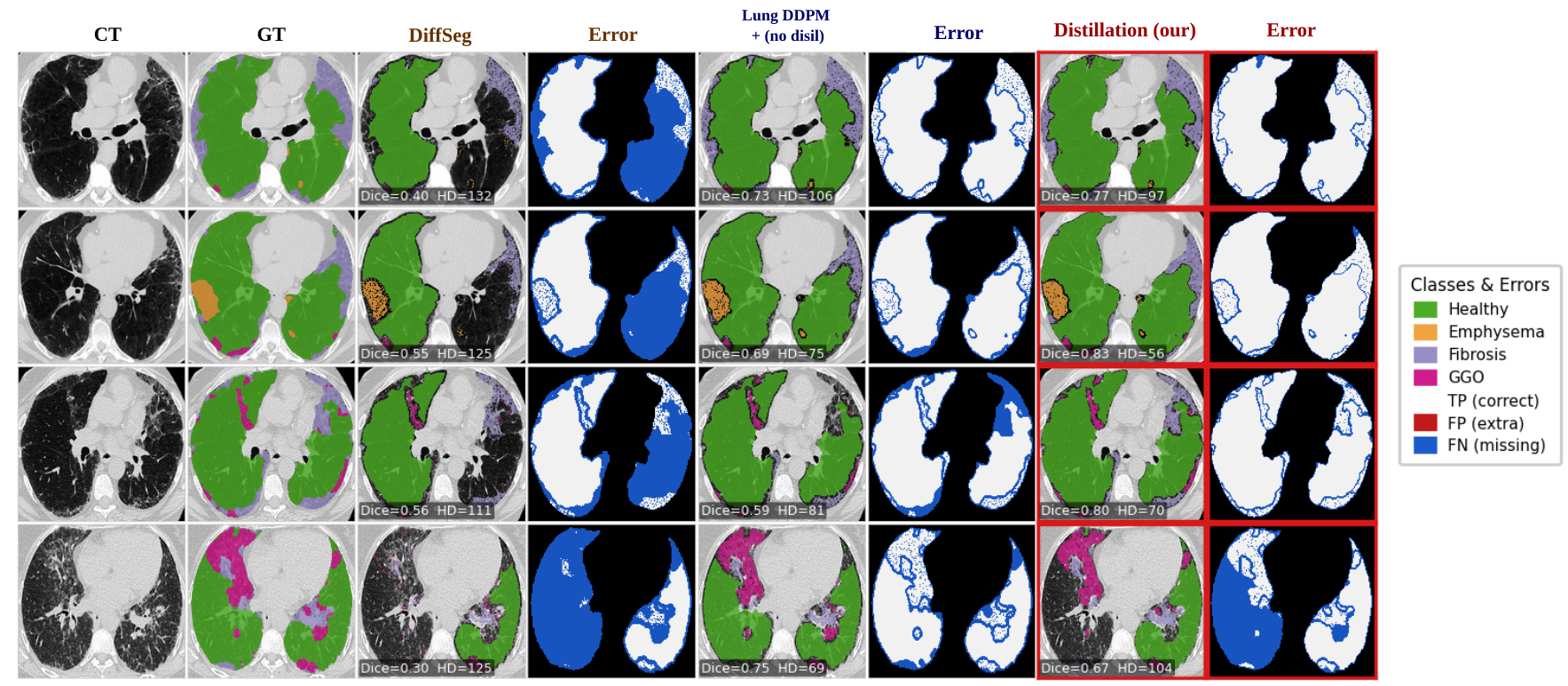}
  \caption{Qualitative segmentation comparison on representative
  axial slices from the 190-slice evaluation set.}
  \label{fig:seg}
\end{figure}

\subsection{Ablation Study}

Table~\ref{tab:ablation} isolates the contribution of each component.
HU preservation yields the largest single DSC gain (+3.5 points),
confirming that physical density is the primary discriminative signal
 particularly for emphysema, whose tissue signature is defined by
its characteristic low HU range. Distillation without warmup regresses
below the HU-only baseline due to bottleneck collapse
(Sec.~\ref{sec:distillation}); adding the warmup schedule recovers
and surpasses it, showing the distillation objective is effective once
the representation is properly initialised. Multi-timestep aggregation
and Sobel-Diffusion Fusion each provide consistent HD95 improvements,
reflecting better boundary localisation. Replacing DPM-Solver with
full DDPM sampling does not affect DSC but increases inference time
by $8\times$, confirming the approximation is lossless for feature
extraction.
\begin{table}[H]
\centering
\caption{Ablation study. Each row enables (\checkmark) or
disables ($\times$) individual components.
DSC and HD95 are averaged over four classes.}
\label{tab:ablation}
\scriptsize
\setlength{\tabcolsep}{2.5pt}
\begin{tabular}{lccccc|cc}
\toprule
& \rotatebox{70}{HU-Preserved}
& \rotatebox{70}{Distillation}
& \rotatebox{70}{Warmup}
& \rotatebox{70}{Multi-timestep}
& \rotatebox{70}{Sobel-Diffusion Fusion}
& DSC\,$\uparrow$
& HD95\,$\downarrow$ \\
\midrule
Baseline (8-bit, no distil.)
  & $\times$ & $\times$ & $\times$ & $\times$ & $\times$ & 71.3 & 18.4 \\
+ HU preservation
  & \checkmark & $\times$ & $\times$ & $\times$ & $\times$ & 74.8 & 16.1 \\
+ Distillation (no warmup)
  & \checkmark & \checkmark & $\times$ & $\times$ & $\times$ & 77.6 & 14.3 \\
+ Warmup schedule
  & \checkmark & \checkmark & \checkmark & $\times$ & $\times$ & 79.9 & 12.7 \\
+ Multi-timestep aggr.
  & \checkmark & \checkmark & \checkmark & \checkmark & $\times$ & 82.1 & 10.5 \\
+ Sobel-Diffusion Fusion
  & \checkmark & \checkmark & \checkmark & \checkmark & \checkmark
  & \textbf{84.0} & \textbf{8.6} \\
\bottomrule
\end{tabular}
\end{table}
\subsection{Conclusion}

We presented DiffSegLung, an unsupervised framework for lung pathology
segmentation that introduces Diffusion Radiomic Distillation:
handcrafted radiomic descriptors serve as a physics-grounded teacher
to shape the bottleneck of a 3D DDPM via a contrastive objective,
without any manual annotations. HU-preserving training is the single
most important component, providing the physical density signal that
supervised methods take for granted. Emphysema remains the most
challenging class due to its HU overlap with healthy parenchyma,
pointing to a natural direction for future work in finer-grained
HU-aware clustering. DiffSegLung simultaneously improves generation
fidelity over prior CT diffusion baselines, demonstrating that
radiomic distillation benefits both discriminative and generative
objectives.


\bibliographystyle{splncs04}
\bibliography{references}

\end{document}